\documentclass[fleqn,twoside]{article}
\usepackage{espcrc2}

\usepackage{graphicx}
\usepackage[figuresright]{rotating}

\newcommand{\AmS}{{\protect\the\textfont2
  A\kern-.1667em\lower.5ex\hbox{M}\kern-.125emS}}

\title{Analyticity in $\theta$ and infinite volume limit of the topological
susceptibility in $SU(3)$ gauge theory}

\author{B. All\'es\address{INFN Sezione di Pisa, 56127--Pisa, Italy},
    M. D'Elia\address{Dipartimento di Fisica dell'Universit{\`a} di
    Genova and INFN, 16146--Genova, Italy},
    A. Di Giacomo\address{Dipartimento di Fisica dell'Universit\`a 
    and INFN, 56127--Pisa, Italy}
}
       
\begin{document}

\begin{abstract}
The large volume behaviour of the topological susceptibility in $SU(3)$ gauge
theory is investigated on the lattice to establish an upper limit on the parity
violating terms.
\vspace{1pc}
\end{abstract}

\maketitle

\section{INTRODUCTION}
\label{sec:introduction}

A theorem by Vafa and Witten~\cite{vafa} states that QCD cannot spontaneously
break parity. However the validity of this result has been questioned by
several authors. At finite temperature there are parity--odd
Lorentz non--invariant operators that may have non vanishing expectation values
in the vacuum~\cite{cohen}. It is also not clear whether the conclusion of the theorem can
be extended to purely fermionic operators~\cite{aoki}. On the other hand
the proof followed by the authors of~\cite{vafa} might implicitly assume the
absence of spontaneous symmetry breaking of parity~\cite{azcoiti}.

Recently Aguado and Asorey~\cite{aguado,asorey} have proved that
parity cannot be broken by the topological charge operator $Q$. The vacuum
energy density $E(\theta)$ is a well defined smooth function for any value
of $\theta$. This excludes the possibility of a cusp at $\theta=0$ and implies
$\langle Q\rangle=0$ and a volume--independent finite value for the topological
susceptibility $\chi$. 

A direct check of this issue on the lattice can be
useful. The idea is to study the volume (in)dependence
of the topological susceptibility $\chi$. One can parametrize the breaking of
parity by writing for the topological charge
\begin{equation}
Q=\pm\alpha V + g\; ,
\end{equation}
where $\alpha$ is a parameter, $V=L^4$ the space--time volume ($L$ is the
lattice size). The quantity $\pm\alpha V$ is the expectation value of
$Q$ in each of the two (possible) broken parity vacua among which, at
finite volume, the system can do tunneling ($\alpha=0$ means no parity
violation). $g$ is a fluctuation whose distribution satisfies
$\langle g\rangle=0$ and $\langle g^2\rangle \neq 0$. Assuming that
$g$ is stochastically independent of the broken parity vacuum, the
topological susceptibility reads
\begin{equation}
\chi = {\langle Q^2 \rangle \over V} = \alpha^2 V +{ \langle g^2\rangle \over V} \;.
\end{equation}
Lattice simulations can give an upper bound to $|\alpha|$ and hence
to the possibility of a spontaneous breaking of parity.

\section{THE LATTICE PROCEDURE}
\label{sec:thelatticeprocedure}

Pure Yang--Mills $SU(3)$ theory was numerically simulated by using the standard
plaquette action on the lattice at $\beta=6.0$ on three different volumes:
$16^4$, $32^4$ and $48^4$. In all cases a heat--bath algorithm was used for
updating configurations and care was taken to decorrelate successive measurements
in order to render them independent. On the lattice we determined the topological
charge with the 1--smeared operator $Q_L^{(1)}$ introduced in~\cite{christou} (see
also~\cite{alles1}). The relationship between the lattice bare topological
susceptibility $\chi_L\equiv \langle\left(Q_L^{(1)}\right)^2\rangle/V$ and
the physical one $\chi$ is given by~\cite{campostrini}
\begin{equation}
\chi_L = a^4 Z^2 \chi + M\; ,
\end{equation}
where $a$ is the lattice spacing, $Z$ and $M$ are renormalization constants
that can be extracted by using a nonperturbative method~\cite{digiacomo,alles2}.
In Eq.(3) $\chi$ is the physical topological susceptibility that coincides
with ${\rm d}^2 E(\theta)/{\rm d}\theta^2$. As it is argued in~\cite{aguado},
if parity is broken a volume dependence of $\chi$ is expected.

In order to provide a clear upper bound to the parity breaking parameter $|\alpha|$,
we have performed high statistics runs for all lattice sizes by using the
APEmille computers in Pisa. The results are shown in Fig.1. The trend 
in this figure is consistent with a volume--independence of $\chi_L$ for large
enough sizes. However, any definite conclusion must be drawn from the physical
topological susceptibility $\chi$.

\begin{figure}[h]
\vspace{9pt}
\includegraphics[scale=0.44]{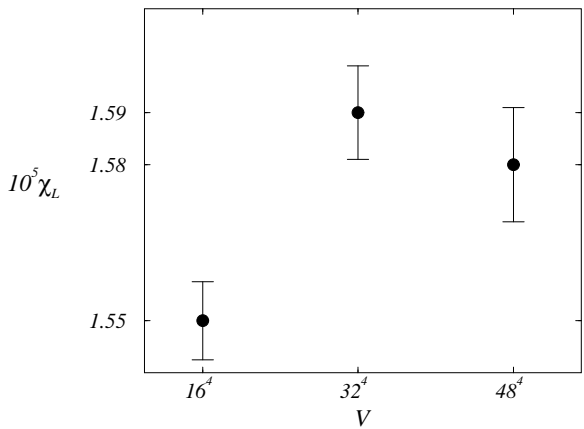}
\caption{Bare topological susceptibility $\chi_L$ at $\beta=6.0$ for
several lattice volumes.}
\label{fig1}
\end{figure}
%

%
%

If the same values of $Z$ and $M$ are used for all volumes, it is clear from Eq.(3)
that Fig.1 will be transformed into an analogous figure that displays the same
trend for the physical topological susceptibility $\chi$. However in perturbation
theory the renormalization constants present a mild dependence on the lattice size
$L$, at least for small $L$. This is a well--known result for several 2--dimensional
models~\cite{hasenfratz}. Although no similar analysis have been carried through
for 4--dimensional QCD, it is expected that in this theory the inclusion of
the zero--modes will show up again as an explicit dependence on $L$.

In principle an $L$ dependence is expected also for the exact nonperturbative
determination of $Z$ and $M$. We are currently calculating
$Z$ and $M$ following the method of~\cite{digiacomo,alles2}. High statistics
is required to reduce the errors. We are using lattice
sizes $8^4$, $12^4$ and $16^4$. The results will be extrapolated by using
the functional forms
\begin{eqnarray}
Z(L)&=& Z_0 + {Z_1 \over L} + {Z_2 \over L^2} \; , \nonumber \\
M(L)&=& M_0 + {M_1 \over L} + {M_2 \over L^2} \; .
\end{eqnarray}


\begin{thebibliography}{9}
\bibitem{vafa} C. Vafa, E. Witten, Phys. Rev. Lett. 53 (1984) 535.
\bibitem{cohen} T. D. Cohen, Phys. Rev. D64 (2001) 047704.
\bibitem{aoki} S. Aoki, Phys. Rev. D30 (1984) 2653; S. Aoki, A. Gocksch,
Phys. Rev. D45 (1992) 3845; S. R. Sharpe, R. J. Singleton, Phys. Rev. D58
(1998) 074501.
\bibitem{azcoiti} V. Azcoiti, A. Galante, Phys. Rev. Lett. 83 (1999) 1518.
\bibitem{aguado} M. Aguado, M. Asorey, hep--th/0204130.
\bibitem{asorey} M. Asorey, Nucl. Phys. Proc. Suppl. 127 (2004) 15.
\bibitem{christou} C. Christou, A. Di Giacomo, H. Panagopoulos, E. Vicari,
Phys. Rev. D53 (1996) 2619.
\bibitem{alles1} B. All\'es, M. D'Elia, A. Di Giacomo, Nucl. Phys. B494
(1997) 281.
\bibitem{campostrini} M. Campostrini, A. Di Giacomo, H. Pana-\break 
gopoulos, E. Vicari, Nucl. Phys. B329 (1990) 683.
\bibitem{digiacomo} A. Di Giacomo, E. Vicari, Phys. Lett. B275 (1992) 429.
\bibitem{alles2} B. All\'es, M. Campostrini, A. Di Giacomo, Y. G\"und\"u\c c,
E. Vicari, Phys. Rev. D48 (1993) 2284.
\bibitem{hasenfratz} P. Hasenfratz, Phys. Lett. B141 (1984) 385.
\end{thebibliography}
\end{document}